\documentclass[aps,prl,twocolumn,showpacs]{revtex4}
\usepackage{amssymb}
\usepackage{amsmath}
\usepackage{graphicx}
\newcommand{\ket}[1]{\left\vert#1\right\rangle}

\begin{document}
\title{Experimental Realization of One-Way Quantum Computing with
Two-Photon Four-Qubit Cluster States}

\author{Kai Chen$^{1,2}$}
\author{Che-Ming Li$^{1,3}$}
\author{Qiang Zhang$^2$}
\author{Yu-Ao Chen$^1$}
\author{Alexander Goebel$^1$}
\author{Shuai Chen$^1$}
\author{Alois Mair$^1$}
\author{Jian-Wei Pan$^{1,2}$}

\affiliation{$^1$Physikalisches Institut,
Ruprecht-Karls-Universit\"{a}t Heidelberg, Philosophenweg 12,
69120 Heidelberg, Germany\\
$^2$Hefei National Laboratory for Physical Sciences at Microscale
and Department of Modern Physics, University of Science and
Technology of China, Hefei, Anhui 230026, China\\
$^3$Department of Electrophysics, National Chiao Tung University,
Hsinchu 30050, Taiwan}

\date{Received 1 May 2007; published 17 September 2007}

\begin{abstract}
We report an experimental realization of one-way quantum computing on a
two-photon four-qubit cluster state. This is accomplished by developing a
two-photon cluster state source entangled both in polarization and spatial
modes. With this special source, we implemented a highly efficient Grover's
search algorithm and high-fidelity two qubits quantum gates. Our experiment
demonstrates that such cluster states could serve as an ideal source and a
building block for rapid and precise optical quantum computation.
\end{abstract}

\maketitle

Highly entangled multipartite states, so-called cluster
states, have recently raised enormous interest in quantum
information processing (QIP). These sorts of states are
crucial as a fundamental resource and a building block
aimed at one-way universal quantum computing \cite{onewayQC}. They
are also essential elements for various quantum error correction
codes and quantum communication protocols \cite{ECC-qcp}.
Moreover, the entanglements are shown to be robust
against decoherence \cite{robust}, and persistent against loss of
qubits \cite{onewayQC}, and thus are exceptionally well suited for quantum
computing and many tasks \cite{onewayQC,ECC-qcp}. Considerable efforts
have been made toward generating and characterizing
cluster state in linear optics \cite{browne-duan,Weinfurter-cluster2005,Zeilinger-clusterbell2005,zeilingernature2005,zeilingernature2007,Lu-Pan2007cluster}.
Recently the principal
feasibility of a one-way quantum computing model has
been experimentally demonstrated through 4-photon cluster
states successfully \cite{zeilingernature2005,zeilingernature2007,onewayDeutsch}.

So far, preparing photonic cluster state still
suffers from several serious limitations. Due to the probabilistic
nature and Poissonian distribution of the parametric down-conversion
process, the generation rate of 4-photon cluster states is quite low
\cite{Zeilinger-clusterbell2005,Weinfurter-cluster2005,zeilingernature2005,zeilingernature2007},
and largely restricts speed of computing. Besides, the quality
and fidelity of prepared cluster states are relatively low
\cite{Weinfurter-cluster2005,zeilingernature2005,zeilingernature2007},
which are difficult to be improved substantially. These
disadvantages consequently impose great challenges of advancement
even for few-qubit quantum computing.

Fortunately, motivated by the progress that an important type of
states termed hyper-entangled states have been experimentally
generated
\cite{AVN2005-martini,AVN2005-pan,Kiwat-hyper2005,Weinfurter-bell2006},
we have the possibility to produce a new type of cluster state
(2-photon 4-qubit cluster state) with nearly perfect fidelity and
high generation rate. The hyper-entangled states have been used to
test ``All-Versus-Nothing" (AVN) quantum nonlocality
\cite{AVN-theory,AVN2005-martini,AVN2005-pan}, and are shown to
lead to an enhancing violation of local realism
\cite{cabello2006,Martini-enhancingbellviolationi2006}. The states
also enable to perform complete deterministic Bell state analysis
\cite{kwiat-weinfurter1998} as demonstrated in
\cite{Weinfurter-bell2006,Martini-bell2006}.

In this Letter we report an experimental realization of one-way
quantum computing with such a 2-photon 4-qubit cluster state. The
key idea is to develop and employ a bright source
which produces a 2-photon state entangled both in polarization and
spatial modes. We are thus able to implement the Grover's algorithm
and quantum gates with excellent performances. The genuine
four-partite entanglement and high fidelity
of better than 88\%
are characterized by an optimal entanglement witness. Inheriting the
intrinsic two-photon
character, our scheme promises a brighter source by 4 orders of
magnitude than the usual
multi-photon source, which offers a significantly high
efficiency for optical quantum computing. It thus provides a simple
and fascinating alternative to complement the latter. With ease
of manipulation and control, the nearly perfect quality of this
source allows to perform highly faithful and precise quantum
computing.

\begin{figure}[ptb]
\begin{center}
\includegraphics[width=8.4cm]{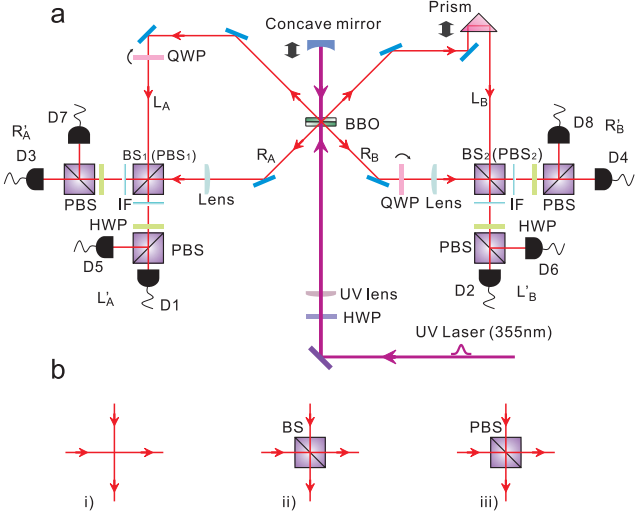}
\caption{Schematic of experimental setup.
\textbf{(a)}. By pumping a two-crystal structured BBO in a double
pass configuration, one polarization entangled photon pair is
generated either in the forward direction or in the backward
direction. The UV pulsed laser (5ps) has a central wavelength of 355
nm with a repetition rate of 80 MHz, and an average power of 200mW.
Two quarter-wave plates (QWPs) are tilted along their optic axis to
vary relative phases between polarization components to attain two
desired possibilities for entangled pair creation. Concave mirror
and prism are mounted on translation stages to optimize interference
on two beam splitters(BS$_{1,2}$) or polarizing beam splitters
(PBS$_{1,2}$) for achieving the target cluster state. Half-wave
plates (HWPs) together with PBS and eight single-photon detectors
(D1-D8) are used for polarization analysis of the output state. IFs
are 3-nm bandpass filters with central wavelength 710 nm.
\textbf{(b)}. In the place where BS$_{1,2}$ or PBS$_{1,2}$ are
located, three apparatuses are for measuring all necessary
observables. Setup \textbf{(i)} is for $Z$ measurement while setup
\textbf{(ii)} is used for $X$ measurement for spatial modes. If an
$\alpha$ phase shifter is inserted at one of the input modes in
(ii), an arbitrary measurement along basis $B(\alpha)$ can be
achieved. Setup \textbf{(iii)} can be for $Z$ measurement of spatial
mode and, simultaneously, for $Z$ measurement of polarization.}
\label{setup}
\end{center}
\end{figure}

To generate the state we use the technique developed
in previous experiments \cite{AVN2005-pan} with type-I spontaneous parametric
down-conversion (SPDC) source \cite{SPDC-I}. The experimental setup
is shown in Fig.~\ref{setup}a. A pulse
of ultraviolet (UV) light passes twice through
two contiguous beta-barium borate (BBO) with optic axes aligned
in perpendicular planes to produce one
polarization entangled photon pair, with one possibility in the forward
direction with a state $(\left\vert H\right\rangle _{A}\left\vert
H\right\rangle _{B}+\left\vert V\right\rangle _{A}\left\vert V\right\rangle
_{B})/\sqrt{2}$ on spatial (path) modes $L_{A,B}$, and another possibility in
the backward direction with a state $(\left\vert H\right\rangle
_{A}\left\vert H\right\rangle _{B}-\left\vert V\right\rangle _{A}\left\vert
V\right\rangle _{B})/\sqrt{2}$ on modes $R_{A,B}$. Here $\left\vert
H\right\rangle $ ($\left\vert V\right\rangle $) stands for photons with
horizontal (vertical) polarization.

Through perfect temporal overlaps of modes $R_{A}$ and $L_{A}$ and of modes
$R_{B}$ and $L_{B}$, one can obtain a state
\begin{eqnarray}
&& \Big(\big(\left\vert H\right\rangle
_{A}\left\vert H\right\rangle _{B}+\left\vert V\right\rangle
_{A}\left\vert V\right\rangle _{B}\big) \left\vert L\right\rangle
_{A}\left\vert L\right\rangle _{B}+ \nonumber\\
&& e^{i\theta}\big(\left\vert H\right\rangle
_{A}\left\vert H\right\rangle _{B}-\left\vert V\right\rangle
_{A}\left\vert V\right\rangle _{B}\big) \left\vert
R\right\rangle _{A}\left\vert R\right\rangle _{B} \Big)/2.
\label{clusterori}
\end{eqnarray}
By properly adjusting the distance between the concave mirror and the crystal
such that $\theta=0$, the generated state will be exactly the desired
cluster state
\begin{eqnarray}
\left\vert C_{4}\right\rangle &=&\big(\left\vert 0000
\right\rangle _{1234}+\left\vert 0011\right\rangle _{1234} \notag \\
&&+\left\vert 1100\right\rangle _{1234}
-\left\vert 1111\right\rangle _{1234}\big)/2,
\label{cluster}
\end{eqnarray}
if we identify the polarization and spatial modes of photon A to be
qubits 2, 3, respectively and photon B's polarization and spatial
modes to be qubits 1,4 and encode logical qubits as $|H(V)\rangle
_{B}\leftrightarrow |0(1)\rangle _{1}$, $|H(V)\rangle
_{A}\leftrightarrow |0(1)\rangle _{2}$, $|L(R) \rangle
_{A}\leftrightarrow |0(1)\rangle _{3}$, $|L(R) \rangle
_{B}\leftrightarrow |0(1)\rangle _{4}$. We observe a cluster state
generation rate about $1.2\times10^4$ per second for 200mW UV pump,
which is 4 order of magnitude brighter than the usual 4-photon
cluster state production
\cite{zeilingernature2005,zeilingernature2007,Weinfurter-cluster2005}
where only a rate of $\sim 1$ is achieved per second.

To evaluate the quality of the state, we apply an optimal
entanglement witness \cite{toth-guehne2005}. The witness is of form
\begin{eqnarray}
\mathcal W&=&\Big(4\cdot I^{\otimes4} -(XXIZ+XXZI+IIZZ \notag \\
&&+IZXX+ZIXX+ZZII)\Big)/2,
\label{witness}
\end{eqnarray}
where $I$ is a 2-dimensional identity matrix while $Z=(|0\rangle \langle
0|-|1\rangle \langle 1|), X=(|0\rangle \langle 1|+|1\rangle \langle 0|)$ are
Pauli matrices. A negative value for the witness implies 4-partite entanglement
for a state close to $\left\vert C_{4}\right\rangle$ and will be optimally as -1 for a
perfect cluster state.
Two experimental settings of $XXZZ$ and $ZZXX$ are needed.
$XXZZ$ can be attained by measuring in the $+/-$ basis for the
polarization in each output arm after apparatus (i) in Fig.~\ref{setup}b.
while $ZZXX$ can be realized by measuring in the $H/V$ basis
after apparatus (ii). This is because the beam splitter (BS) acts exactly as a Hadamard
transformation for the path modes to change $Z$ basis to $X$ basis for
measurement, namely, $\left\vert L\right\rangle_{A,B}\rightarrow (\left\vert
R'\right\rangle _{A,B} +\left\vert L'\right\rangle _{A,B})/\sqrt{2}$,
$\left\vert R\right\rangle _{A,B}\rightarrow (\left\vert R'\right\rangle _{A,B}
-\left\vert L'\right\rangle _{A,B})/\sqrt{2}$. All of the observables for
evaluating the witness are listed in Table \ref{witnessmeasurent}.
\begin{table}[t]
\begin{tabular}{ccccc}
\hline\hline Observable &Value &Observable
&Value\\
\hline
$XXIZ$& $0.9070\pm0.0036$ & $IZXX$ &  $0.9071\pm0.0037$&\\
$XXZI$& $0.9076\pm0.0035$ & $ZIXX$ &  $0.8911\pm0.0040$&\\
$IIZZ$& $0.9812\pm0.0016$ & $ZZII$ &  $0.9372\pm0.0030$&\\
\hline\hline
\end{tabular}
\caption{Experimental values of all the observable on the
state $|C_{4}\rangle$ for the entanglement witness $\mathcal W$
measurement. Each experimental value corresponds to measure in an
average time of 1 sec and considers the Poissonian counting
statistics of the raw detection events for the experimental errors.}
\label{witnessmeasurent}
\end{table}
Substituting their experimental values into
Eq.~(\ref{witness}) yields $\langle \mathcal W
\rangle_{exp}=-0.766\pm0.004$, which clearly proves the genuine
four-partite entanglement by about $200$ standard deviations. As shown in
\cite{toth-guehne2005}, one can obtain a lower bound for fidelity of
experimental prepared state to $\left\vert C_{4}\right\rangle$
\begin{eqnarray}
F&\geq&\frac{1}{2}-\frac{1}{2}\langle \mathcal W \rangle_{exp}=0.883\pm0.002.
\label{fidelity}
\end{eqnarray}
This proves to be a better
source than the ones in
\cite{zeilingernature2005,zeilingernature2007,Weinfurter-cluster2005}
where fidelities are about 0.63
\cite{zeilingernature2005,zeilingernature2007} and 0.74
\cite{Weinfurter-cluster2005} respectively. We attribute impurity of
our state to imperfect overlapping on BS, deviations of BS from
50\%, as well as imperfections in the polarization and path modes
analysis devices. To get a qualitative depiction for these
imperfections, we scan the concave mirror with nanometer
displacements and observe interference after BS$_{1,2}$. By
measuring along $H/V$ basis in each output arm, we have obtained
visibility of $0.842\pm 0.008, 0.943\pm 0.006, 0.968\pm 0.004,
0.949\pm 0.006$ for coincidences among detectors D1-D2,D1-D4,D3-D2
and D3-D4 respectively.

\begin{figure}[t]
\includegraphics[width=8.4cm]{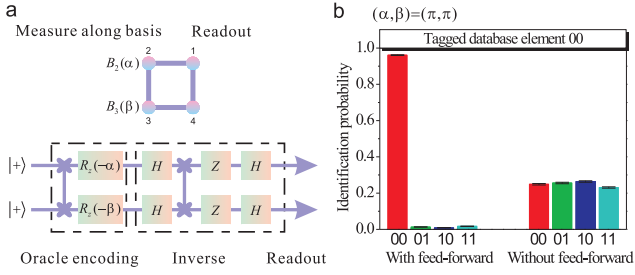}
\caption{Demonstration of Grover's algorithm.
\textbf{a}. Equivalent quantum circuit of
Grover's algorithm using box cluster state. The `oracle' encodes the
element `00' by measuring along basis $B_{2,3}(\pi)$, while the
inverse and readout sections will find this entry with certainty by
a single query. \textbf{b}. A successful identification probability
of $(96.1\pm0.2)\%$ is achieved deterministically with feed-forward,
while it is $(24.9\pm0.4)\%$ without feed-forward. This is in
an excellent agreement with theoretical expectations.
The trick is that the black box provides only outcomes
but not basis information for feed-forward. Thus the oracle encoding
is hidden before feed-forward on readout.}
\label{grover}
\end{figure}

A cluster state can be represented by an array of nodes, where
each node is initially in the state of $\left|+\right\rangle=
\left( \left| 0\right\rangle +\left| 1\right\rangle
\right)/\sqrt{2}$. Every connected line between nodes experiences
a controlled-phase (CPhase) gates acting as $\left|
j\right\rangle \left| k\right\rangle \rightarrow \left( -1\right)
^{jk}\left| j\right\rangle \left| k\right\rangle ,$ $j,k\in
\left\{ 0,1\right\}$ \cite{onewayQC}. For a given cluster state,
consecutive single-qubit measurements in basis $B_k(\alpha)=\{
\ket{\alpha_+}_k,\ket{\alpha_-}_k \}$ will define a quantum
computing in addition to feed-forward of measurement outputs,
where $\ket{\alpha_{\pm}}_k=(\ket{0}\pm e^{i
\alpha}\ket{1})_k/\sqrt{2}$ ($\alpha\!\in\!{\mathbb R}$). A
measurement output of $\ket{\alpha_+}_k$ means `0' while
$\ket{\alpha_-}_k$ signifies `1'. This measurement basis
determines a rotation $R_z(\alpha)={\rm exp}(-i \alpha Z/2)$,
followed by a Hadamard operation $H=(X+Z)/\sqrt{2}$ of encoded
qubits. The state $\ket{C_{4}}$ can be represented by a box type
graph shown in Fig.~\ref{grover}a, up to a local unitary
transformation.

{\em Grover's algorithm.} For an unsorted database with $N$ entries,
Grover's search algorithm gives a quadratic speed-up for with
$\sim\sqrt{N}$ consultations on average \cite{groveralgorithm}.
Striking linear optics implementations have been achieved in
\cite{kwiat-groveralgorithm,Groverclassical}, although it is
questionable whether the algorithm is truly `quantum' due to a
demonstration \cite{Groverclassical} based on interference of
classical waves. One-way realizations have been carried out
\cite{zeilingernature2005,zeilingernature2007} recently. In the case
of four entries $\ket{00},\ket{01},\ket{10},\ket{11}$, a single
quantum search will find the marked element  An execution goes as
follows: an oracle encodes a desired entry by changing its sign
through a black box with initial state $\ket{++}$. After an
inversion-about-the-mean operation, the labeled element will be
found with certainty by readout. It is shown in
\cite{zeilingernature2005} that this can be exactly finished with
the box cluster state in Fig.~\ref{grover}a. For demonstration, we
experimentally tag the element $\ket{00}$ on qubits $2,3$ and make
the readout on qubits $1,4$ all along basis $B(\pi)$. Because of the
fact that the state Eq.~(\ref{cluster}) distinguishes the box
cluster from an $H$ transformation on every qubit and a swap between
qubits 2 and 3, this amounts to measuring along the $V/H$ basis
after apparatus (iii) in Fig.~\ref{setup}b. Two polarizing beam
splitters (PBS) here are for interfering, to ensure the desired
cluster state. In the meantime they are acting as polarization
measurement devices, which is equivalent to use apparatus (i) in
this case. The outputs of the algorithm are two bits $\{s_3\oplus
s_4,s_1\oplus s_2\}$ in lab basis by feed-forwarding outcomes of
qubits 2,3, where $s_i$ are measurement outcomes on qubits $i$. The
experimental results are sketched in Fig.~\ref{grover}b.
\begin{figure}[t]
\includegraphics[width=7.3cm]{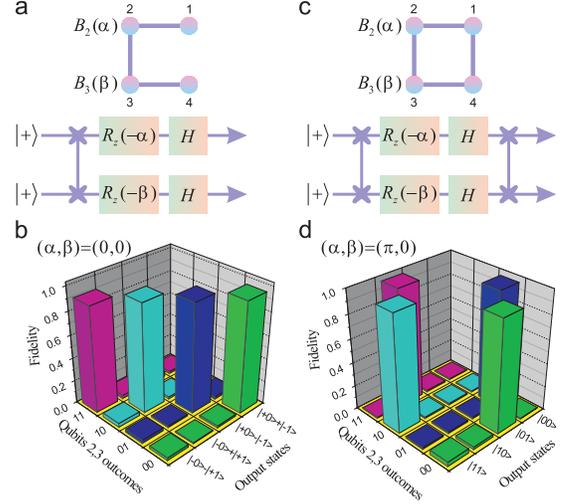}
\caption{Two-qubit quantum gates realizations. \textbf{a}. CPhase
gate realization with the horseshoe cluster. \textbf{b}.
Experimental measured fidelities of output states to the ideal Bell
states (unnormalized) in the lab basis. They are
$0.954\pm0.003,0.940\pm0.004,0.936\pm0.005,0.910\pm0.005$ for
outcomes 00,01,10,11 on qubits 2,3 respectively.
\textbf{c}. Quantum gate implementation that does not generate
entanglement with the box cluster. \textbf{d}.
Measured fidelities of output states to the
ideal product states in the lab basis. They are
$0.935\pm0.005,0.962\pm0.004,0.969\pm0.003,0.975\pm0.003$ for
outcomes 00,01,10,11 on qubits 2,3 respectively.}
\label{q-gates}
\end{figure}

{\em Quantum gates.} Non-trivial two-qubit quantum gates such as the
CPhase gate are at the heart of universal quantum computation,
that can be realized by cluster
states conveniently. Depending on the initial cluster state and
measurement basis, states with
different degrees of entanglement can be generated. The
horseshoe or box cluster shown in Fig.~\ref{q-gates}a and \ref{q-gates}c can
realize such important gates. For the case of horseshoe cluster in
Fig.~\ref{q-gates}a, depending on the outcomes when measuring
along basis $B_{2}(\alpha)$ and $B_{3}(\beta)$, the output state on qubits 1,4
would be $\ket{\Omega_{out}}=(X^{s_2}\otimes X^{s_3})(H\otimes
H)\big(R_z (-\alpha) \otimes R_z (-\beta)\big) \text{CPhase}
\ket{\Omega_{in}}$ where $\ket{\Omega_{in}}=\ket{++}$.
The state $\ket{\Omega_{out}}$ is always a maximal entangled state. Taking
$\alpha=\beta=0$ and consider
outcomes `00' in qubits 2,3. This implies a final Bell state of
$\ket{\Omega_{out}}=(\ket{+}\ket{0}+\ket{-}\ket{1})/\sqrt{2}$. Note
that the horseshoe cluster state is equivalent to the state
Eq.~(\ref{cluster}) up to a $HIIH$ transformation, in lab basis
this amounts to the fact that the output state is exactly
$\ket{\Omega_{out}}$, that is symmetric under $HH$ transformation.
To characterize quality of quantum gates outputs,
we put a birefringent crystal in path $R_B$ to
make a transformation $\ket{+}\leftrightarrow\ket{-}$ for
polarization. After BS$_2$, all the Bell states
on qubits 1,4 will change as
\begin{equation}
\begin{array}{l}
(\ket{+}_1\ket{0}_4\pm\ket{-}_1\ket{1}_4)/\sqrt{2}\longrightarrow\ket{+}_1\ket{\pm}_4, \\
(\ket{-}_1\ket{0}_4\pm\ket{+}_1\ket{1}_4)/\sqrt{2}\longrightarrow\ket{-}_1\ket{\pm}_4,
\end{array}
\label{discrimination}
\end{equation}
which can be completely and deterministically discriminated
by measuring along $\ket{\pm}$ basis.
The fidelities of the output
states in the lab basis to the ideal Bell state are shown in
Fig.~\ref{q-gates}b. Similarly, for the box cluster state shown in
Fig.~\ref{q-gates}c, measurements on qubits 2,3
along basis $\{B_2 (\alpha),B_3 (\beta)\}$
will give an output state on qubits 1,4 with $\ket{\Omega_{out}}=(Z\otimes
X)^{s_3}$$(X\otimes Z)^{s_2}$\text{CPhase}$(H\otimes H)$$\big(R_z (-\alpha)
\otimes R_z (-\beta)\big)$\text{CPhase}$\ket{\Omega_{in}}$ which is a product
state when $\alpha=\pi$ and $\beta=0$. Since we can completely distinguish 4
different products states, output fidelities
can be obtained directly, as shown in Fig.~\ref{q-gates}d. By employing
the techniques developed in \cite{zeilingernature2007} with active
feed-forward, one can expect to achieve deterministically quantum computing
with excellent quality outputs.

We remark that other 2-qubit states can be generated, by
suitable measurements on qubits 2,3.
However, an arbitrary single-qubit rotation needs generally 3
single-qubit measurements on a cluster for one-way
implementation \cite{zeilingernature2005,zeilingernature2007}, which
is a big consuming of resource.
Fortunately, this rotation can be easily attained by linear optical
components both for polarization and spatial modes. Therefore a
hybrid framework would be more practical
with one-way realization of two-qubit gates and the usual
single-qubit gates.
Due to low efficiency for producing multi-photon and concurrent
occupations for polarization-spatial degrees of freedom of the
photons, our source is not yet scalable,
the same as the multi-photon source \cite{zeilingernature2007}. However,
the scheme developed here leads to quantum computing with a
quality and efficiency at present largely unmatched by previous
methods.

In summary, we have developed a scheme for preparation of a
2-photon 4-qubit cluster state, designed and demonstrated the
first proof-of-principle realization of one-way quantum computing
employing such a source. The excellent quality of the state with
fidelity better than $88\%$ is achieved. The high count rates
enable quantum computing by 4 orders of magnitude more efficient
than previous methods. We have implemented the Grover's algorithm
with a successful probability of about $96\%$ and quantum gates
with high fidelities of about $95\%$ on average. Our scheme helps
to make a significant advancement of QIP, and the source
constitutes a promising candidate for efficient and high quality
one-way optical quantum computing. By using more photons and more
degrees of freedom, one can expand our ability to generate
many-qubit cluster states for performing quantum computing and
other complex tasks. Our results can also find rapid applications
in quantum error correction codes, multi-partite quantum
communication protocols \cite{ECC-qcp}, as well as novel types of AVN
tests for nonlocality \cite{AVN-theory}.

This work was supported by the Marie Curie Excellence Grant of the
EU, the Alexander von Humboldt Foundation, the CAS and
the National Fundamental Research Program (Grant No. 2006CB921900).

\textit{Note added}.-- During preparation of our
manuscript, we are aware of one related experiment for realization
of a linear cluster state \cite{Martiniprl2007}.

\end{document}